\documentclass[letterpaper,12pt]{article}
\usepackage[letterpaper, top=2cm, bottom=2.5cm, left=2cm, right=2cm]{geometry}

\parskip=\baselineskip

\usepackage{amssymb,amsmath}
\usepackage[english]{babel}
\usepackage{latexsym}
\usepackage{hyperref}

\newcommand{\Theo}[2]{\begin{description}\item[\textbf{Theorem:}]\,(\emph{#1}) #2\end{description}}

\newcommand{\dd}{{\rm d}}

\newcommand{\equ}{\begin{eqnarray}}
\newcommand{\fequ}{\end{eqnarray}}
\newcommand{\eqn}{\begin{eqnarray}}
\newcommand{\feqn}{\end{eqnarray}}

\newcommand{\M}{{\cal M}}

\newcommand{\F}{{\cal F}}
\newcommand{\A}{{\cal A}}

\begin{document}
\begin{titlepage}
\begin{center}
$\phantom{1}$

\vspace{1.5cm}

{\textbf{\Large Gravity in the 3+1-Split Formalism I:\\}}
\vspace{.1cm}
{\textbf{\Large Holography as an Initial Value Problem}}

\vspace{1cm}

{\large D.~S.~Mansi\footnote{\href{mailto:dmansi@physics.uoc.gr}{\tt dmansi@physics.uoc.gr}} and A.~C.~Petkou\footnote{\href{mailto:petkou@physics.uoc.gr}{\tt petkou@physics.uoc.gr}}}

\vspace{0.2cm}

{\it Department of Physics, University of Crete, GR-71003 Heraklion.}

\vspace{0.5cm}

{\large G.~Tagliabue\footnote{\href{mailto:giovanni.tagliabue@mi.infn.it}{\tt giovanni.tagliabue@mi.infn.it}}}

\vspace{0.2cm}

{\it Dipartimento di Fisica dell'Universit\`a di Milano, Via Celoria 16, I-20133 Milano\\
and INFN, Sezione di Milano, Via Celoria 16, I-20133 Milano.}
\end{center}

\vspace{.5cm}

\begin{abstract}
We present a detailed analysis of the 3+1-split formalism of gravity in the presence of a cosmological constant. The formalism helps revealing the intimate connection between holography and the initial value formulation of gravity. We show that the various methods of holographic subtraction of divergences correspond just to different transformations  of the canonical variables, such that the initial value problem is properly set up at the boundary. The renormalized boundary energy momentum tensor is a component of the Weyl tensor.
\end{abstract}

\end{titlepage}
\parskip=.1\baselineskip
\tableofcontents
\parskip=.3\baselineskip

\section{Introduction}
An enormous amount of work in the past decade has been devoted to  holographic studies of four-dimensional quantum field theories. The basic setup has been the AdS$_5$/CFT$_4$ correspondence, whereby various  five or higher-dimensional gravity models (related or not to supergravity) provide information for strongly coupled, largely supersymmetric, four-dimensional gauge theories. This way, fundamental quantum field theory properties such as symmetry breaking, confinement and finite-temperature phase transitions may be viewed as the holographic images of certain properties of gravitational theories. 

However, relatively less work has been devoted to holographic studies of four-dimensional gravity theories, mainly due to the lack of understanding of their three-dimensional boundary counterparts. Nevertheless, recently there is a sharp rise in interest on AdS$_4$/CFT$_3$ holography. This is partly due to the set of ideas regarding the holographic description of three-dimensional condensed matter systems, see for example \cite{Sachdev,HP,HH,Horowitz, Gubser} and references therein.  Moreover, important additional motivation to study the AdS$_4$/CFT$_3$ correspondence comes from the recent emergence of various proposals regarding three-dimensional theories that describe M2 branes \cite{BL1,BL2,BL3,Gustavsson,ABJM}. 

Apart from attracting all that recent recent interest, AdS$_4$/CFT$_3$ has for some time now being singled out as a new holographic paradigm as it possesses some special properties not shared by  its more known AdS$_5$/CFT$_4$ counterpart. Its most distinctive property is that it gives rise to a  holographic map of the electric-magnetic duality of Yang-Mills,  and also of the {\it generalized electric-magnetic duality} of linearized gravity and higher-spin gauge fields in four dimensions. Some of the salient features of the three-dimensional boundary systems, such as Quantum-Hall type of dualities \cite{BD,Witten,LP1} and the possibility of an exact holography \cite{dHP1,dHG,dHPP} are intimately connected to it \cite{dHPConf,Marolf}.  

We believe that all the above is strong motivation to revisit four-dimensional gravity with a cosmological constant and analyze in depth its holographic description. We embark in this endeavour in the present and a companion work  \cite{MPT2} based on the  3+1-split formalism of \cite{LP2}. The latter was instrumental in the proof \cite{LP2} of electric-magnetic duality of linearized gravity in (A)dS$_4$. We believe that the economy and the familiar physics picture drawn by the 3+1-split formalism (e.g. the introduction of "electric" and "magnetic" gravitational fields)  make it very well suited for studies in the AdS$_4$/CFT$_3$ correspondence. 

The 3+1-split formalism is a hybrid of the standard ADM construction \cite{ADM} and hence it can be used to setup an initial value formalism for gravity. We should clarify from the beginning, however, that the  initial value formalism relevant to  holography is physically different - and in many cases simpler since causality issues do not arise - from the standard initial value formalism that evolves the data on a Cauchy surface along real time. We comment further in Section 3. Perhaps expectedly, we find that in the presence of a cosmological constant, setting up the initial value problem in the boundary is equivalent to a holographic description.\footnote{Throughout the present and the companion work \cite{MPT2} ``holography'' is a broader notion (i.e. the mapping of generic bulk data to the boundary) while AdS/CFT has a more specific meaning (the holographic mapping between specified bulk and boundary theories). Away from a string/M-theory setup it is not clear if generic bulk gravitational data can be encoded by non-gravitational QFTs, nevertheless the issue is not yet settled.}  The upshot of our work is the demonstration that all known methods of holographic removal of divergences, namely holographic renormalization \cite{BK,dHSS} and Kounterterms \cite{Zanelli,Olea} correspond to just different transformations of the canonical variables, such that the initial value problem is properly set up at the boundary. These transformations are canonical when they are implemented on the restricted phase space defined by constraints. We identify the {\it initial boundary velocity} with a component of the Weyl tensor. Holographically the latter gives the boundary energy momentum tensor. In the companion paper \cite{MPT2} we  discuss the notion of self-duality in gravity with a cosmological constant and show the relevance of the three-dimensional gravitational Chern-Simons theory for self-dual configurations. 

We begin in section 2 with a detailed presentation of the the 3+1-split formalism for gravity with Lorentzian signature. We define our variables and explain our gauge-fixing choices. We end up with a compact form of the equations of motion and zero-torsion constraints. Section 3 contain the formal setup of the initial value problem for four-dimensional gravity with a cosmological constant.  In Section 4 we detail the Fefferman-Graham expansion in the 3+1-split formalism. We identify the proper boundary data, namely the {\it initial position} and {\it initial velocity}. We also note that the various terms on the FG expansion correspond to boundary geometrical data. 
In section 5 we explain why holography can be viewed as an initial value formulation of gravity in the boundary. We show that the two different methods of holographic removal of divergences correspond to certain transformations of he canonical variables, such that a proper initial value problem is setup at the boundary. We conclude in Section 6. Three Appendices contain useful relations for the Weyl tensor, a brief presentation of the fist-order formalism for Yang-Mills theories and also the holographic description of Schwarzschild and Taub-NUT AdS black holes.

\section{Details on the the 3+1-split formalism}
In this section we present a concise version of the 3+1-split formalism of \cite{LP2} for  gravity in the presence of non-zero cosmological constant. We consider a Lorentzian manifold $\M$ and take the Einstein-Hilbert action with cosmological constant in the first-order Palatini formalism as
\equ\label{EHP}
S_{{\rm EH}}=-\frac1{32\pi G}\int_{\M}\epsilon_{abcd}\left(R^{ab}+\frac\Lambda2e^a\wedge e^b\right)\wedge e^c\wedge e^d\,.
\fequ
This is thus equivalent to the standard second-order gravitational action 
\begin{eqnarray*}
S_{{\rm2^{nd}}}=-\frac1{16\pi G}\int\dd^{4}x\sqrt{-g}\left(R+6\Lambda\right)\,,
\end{eqnarray*}
and hence the cosmological constants is related to the parameter $\Lambda$ as $\Lambda_{\rm cosm.}=-3\Lambda$.
The curvature and torsion 2-forms are defined in terms of the vielbein $e^a$  and spin-connection $\omega^{ab}$ as
\begin{eqnarray*}
R^{ab}=\dd\omega^{ab}+\omega^a{}_{c}\wedge\omega^{cb}\,,\qquad T^a=\dd e^a+\omega^a{}_{b}\wedge e^b\,.
\end{eqnarray*}
We define $\eta_{ab}={\rm diag}(\sigma_{\perp},+,+,\sigma_3)$, where $\sigma_{\perp}\sigma_3=-1$, $\sigma_\perp^2=\sigma_3^2=1$ and set $\Lambda=\sigma_{\perp}/\ell^{2}$ such that $\Lambda <0$ ($\Lambda >0$) yields the de Sitter (Anti-de Sitter) vacuum. The manifold is supposed it can be foliated by slices $\Sigma_{t}$ indexed by a function $t$ which is either a time coordinate if $\sigma_{\perp}=-1$ or a radial coordinate if $\sigma_{\perp}=+1$. Consequently, we split the vielbein and the spin connection as\footnote{Throughout this work, Latin indices run as $a,b,c...=0,1,2,3$ and Greek indices as $\alpha,\beta,\gamma,...=1,2,3$.} 
\eqn
e^0&=&N\dd t\,,\qquad e^\alpha=N^\alpha\dd t+\tilde e^\alpha\,,\label{esplit}\\
\omega^{0\alpha}&=&q^{0\alpha}\dd t+\sigma_{\perp}K^\alpha\,,\qquad\omega^{\alpha\beta}=-\epsilon^{\alpha\beta\gamma}\left(Q_\gamma\dd t+B_\gamma\right)\,.\label{omegasplit}
\feqn
Using (\ref{esplit}) and (\ref{omegasplit}) and some lengthly but straightforward calculations \cite{LP2} the action (\ref{EHP}) can be brought into the form 
\eqn
S_{{\rm EH}}&=&-\frac{\sigma_{\perp}}{8\pi G}\int_{\M}\dd t\wedge\left\{-K_{\alpha}\wedge\dot\Sigma^\alpha+N\tilde W_{\alpha}\wedge\tilde e^{\alpha}+\sigma_\perp\hat Q\wedge K_{\beta}\wedge\tilde e^{\beta}\right.\nonumber\\
&&\phantom{-\frac{\sigma_{\perp}}{8\pi G}\int_{\M}\dd t\wedge\left\{\right.}\left.+\sigma_{\perp}q^{0\alpha}\tilde{\cal D}\Sigma_{\alpha}-N^{\alpha}\epsilon_{\alpha\beta\gamma}\tilde{\cal D}K^{\beta}\wedge\tilde e^{\gamma}\right\}\nonumber\\
&&-\frac1{8\pi G}\int_{\partial\M}\left(q^{0\alpha}\dd t+\sigma_{\perp}K^{\alpha}\right)\wedge\Sigma_{\alpha}\,,\label{lor.act}
\feqn
where $\hat Q\equiv Q_{\alpha}\tilde e^{\alpha}$. We have introduced the 2-form
\equ
\tilde W_{\alpha}\equiv\rho_{\alpha}-\frac12\epsilon_{\alpha\beta\gamma}K^{\beta}\wedge K^{\gamma}+\frac1{\ell^{2}}\Sigma_{\alpha}\,.\label{tildeW}
\fequ
and have defined the oriented surface element as
\begin{eqnarray*}
\Sigma^{\alpha}={}^{\tilde*}\tilde e^{\alpha}=\frac12\epsilon^{\alpha}{}_{\beta\gamma}\tilde e^{\beta}\wedge\tilde e^{\gamma}\,,
\end{eqnarray*}
with ${}^{\tilde*}$ the three-dimensional Hodge dual defined in terms of $\tilde e^\alpha$ only. The three-dimensional component of the curvature 2-form 
\begin{eqnarray*}
\rho_{\alpha}=\tilde\dd B_{\alpha}+\frac12\epsilon_{\alpha\beta\gamma}B^{\beta}\wedge B^{\gamma}\,,
\end{eqnarray*}
is made out of $B^\alpha$ only. Moreover, $\tilde{\cal D}$ denotes a covariant derivative with respect to the one-form field $B^{\alpha}$ as 
\begin{eqnarray*}
\tilde{\cal D}V^\alpha =\tilde\dd V^\alpha +\epsilon^{\alpha}_{\,\,\beta\gamma}B^\beta\wedge V^\gamma\,,
\end{eqnarray*}
if $V^\alpha$ is a generic vector-valued one-form (with respect to either ${\rm SO}(3)$ or ${\rm SO}(2,1)$ depending on whether $\sigma_\perp=\mp1$ respectively) defined on $\Sigma_t$. Comparing the action (\ref{lor.act}) to the Yang-Mills action \eqref{YM-action} in \ref{Lor.YM}  motivates calling the vector-valued  one-forms $K^\alpha$ and $B^\alpha$ the ``electric" and ``magnetic"  fields respectively. 

The boundary term in (\ref{lor.act}) is exactly {\it minus} the usual  Gibbons-Hawking term \cite{Gibbons:1976ue} $S_{{\rm GH}}$. Hence, the action $S=S_{{\rm EH}}+S_{{\rm GH}}$ is stationary on-shell when $\delta\tilde{e}^\alpha=0$ in the boundary, i.~e.~it provides a good Dirichlet variational principle with respect to the vielbein. The form of the action (\ref{lor.act}) appears to indicate that the proper conjugate dynamical variables are $\Sigma^{\alpha}$ (or, equivalently, $\tilde e^\alpha$) and $K^\alpha$. We will later see that the proper identification of the dynamical variables is slightly more involved than this.  The remaining fields  $\{N,N^\alpha,q^{0\alpha},\hat Q,B^{\alpha}\}$ enter the action as Lagrange multipliers of the following constraints:
\eqn
-8\pi G\sigma_{\perp}\frac{\delta S}{\delta N}&=&\tilde W_{\alpha}\wedge\tilde e^{\alpha}=0\,,\label{lC.1}\\
-8\pi G\sigma_{\perp}\frac{\delta S}{\delta N^\alpha}&=&-\epsilon_{\alpha\beta\gamma}\tilde{\cal D}K^\beta\wedge\tilde e^\gamma=0\,,\label{lC.2}\\
-8\pi G\sigma_{\perp}\frac{\delta S}{\delta q^{0\alpha}}&=&\sigma_{\perp}\tilde{\cal D}\Sigma_{\alpha}=\sigma_{\perp}\epsilon_{\alpha\beta\gamma}\tilde T^{\beta}\wedge\tilde e^{\gamma}=0\,,\label{lC.3}\\
-8\pi G\sigma_{\perp}\frac{\delta S}{\delta\hat Q}&=&\sigma_\perp K_\alpha\wedge\tilde e^\alpha=0\,,\label{lC.4}\\
-8\pi G\sigma_{\perp}\frac{\delta S}{\delta B^{\alpha}}&=&N\tilde T^{\alpha}+\left(\tilde\dd N+\sigma_{\perp}K_{\beta}N^{\beta}-\hat q\right)\wedge\tilde e^{\alpha}=0\,, \label{lC.5}
\feqn
where $\hat q\equiv q^{0}{}_{\alpha}\tilde e^{\alpha}$. The exterior multiplication of (\ref{lC.5}) by $\epsilon_{\alpha\beta\gamma}\tilde{e}^\gamma$ gives, by virtue of (\ref{lC.3}),
\equ
\tilde\dd N+\sigma_{\perp}K_{\beta}N^{\beta}-\hat q=0\,,\label{lC.6}
\fequ
and hence we obtain the zero torsion condition
\equ
\label{torsion0}
\tilde T^{\alpha}={\cal D}\tilde{e}^\alpha=0\,,
\fequ
In other words, the magnetic field $B^\alpha$ is a Lagrange multiplier which is algebraically related to the vielbein via the vanishing of torsion (\ref{torsion0}). This is exactly analogous to electromagnetism and gives an important hint regarding the relevance of torsion to holography and gravitational duality \cite{LNP}. 
  
Next, we use diffeomorphisms and local Lorentz rotations to fix some of the Lagrange multipliers. 
\begin{description}
\item[{\bf $\{N,N^{\alpha},q^{0\alpha}\}$-fixing}:] Using suitable diffeomorphisms we can fix $N=1$ and $N^\alpha=0$. In order to set $N=1$ it is sufficient to choose, in a certain neibourhood, the proper ``time''\footnote{The coordinate $t$ is actually a time coordinate only if $\Lambda\leq0$. Otherwise it is a spatial coordinate, but the arguments we give above do not change.} of a family of ``timelike'' geodesics as the new time coordinate. Moreover, in order to have $N^\alpha=0$, it is sufficient to choose as new spatial coordinates the coordinates that parametrize the surfaces orthogonal to the family of geodesics we have chosen above. All that means that the spacetime metric can be  cast into the Gaussian normal form
\begin{eqnarray*}
\dd s^{2}=\sigma_{\perp}\dd t^{2}+h_{ij}(t,\vec x)\dd x^{i}\dd x^{j}\,,
\end{eqnarray*}
which is also suitable for discussing holography.
If we use now \eqref{lC.6} we obtain that $q^{0\alpha}=0$.

\item[{\bf $\{Q^{\alpha}\}$-fixing:}] These can be fixed by a suitable local Lorentz rotation. Recall that $\omega$ is an ${\rm so}(3,1)$-valued connection, while the vielbein $e$ is a vector under ${\rm SO}(3,1)$-rotations. Under a generic finite local Lorentz transformation $g\in{\rm SO}(3,1)$ they transform as
\begin{eqnarray*}
e\;&\mapsto&\;e'=g e\,,\\
\omega\;&\mapsto&\;\omega'=g\omega g^{-1}+g\dd g^{-1}\,.
\end{eqnarray*}
If we want to preserve our choice of the vielbein, say $e^{0}=Ndt$, it turns out that $g^{0}{}_{\alpha}=0$. As a consequence we restrict our interest to the subgroup of local trasformations  given by
\begin{eqnarray*}
{\rm L}=\left\{\begin{array}{lcl}{\rm SO}(3) &&{\rm if}\;\,\sigma_\perp=-1\\
{\rm SO}(2,1) &&{\rm if}\;\,\sigma_\perp=+1\end{array}\right\} \subset{\rm SO}(3,1)\,.
\end{eqnarray*}
Then, from the second equation in (\ref{omegasplit}) we see that $Q_\alpha$ can be gauged fixed to zero by a suitable $g\in{\rm L}$ such that 
\begin{eqnarray*}
-\epsilon^{\alpha}{}_{\beta\gamma}Q^{\gamma}=(g^{-1})^{\alpha}{}_{\gamma}\dot g^{\gamma}{}_{\beta}\,.
\end{eqnarray*}
A residual gauge freedom, $t$-independent $L$-rotations on the fields, remains nevertheless. 
\end{description}

With the $N=1$ and $N^\alpha=q^{0\alpha}=Q^\alpha=0$ gauge fixing the equations of motion read
\eqn
-8\pi G\sigma_{\perp}\frac{\delta S}{\delta K^\alpha}&=&-\epsilon_{\alpha\beta\gamma}\left(\dot{\tilde e}^\beta+K^\beta\right)\wedge\tilde e^\gamma=0\,,\label{lE.1}\\
-8\pi G\sigma_{\perp}\frac{\delta S}{\delta \tilde e^\alpha}&=&\tilde W_{\alpha}+\frac2{\ell^{2}}\Sigma_{\alpha}+\epsilon_{\alpha\beta\gamma}\tilde e^\beta\wedge\dot K^\gamma=0\,.\label{lE.2}
\feqn
\eqref{lE.1} actually implies that
\begin{eqnarray*}
\dot{\tilde e}^\alpha+K^\alpha=0\,.
\end{eqnarray*}
Gathering all together,  the equations describing any classical gravitational background in 4D are; the zero torsion conditions 
\equ
K_\alpha\wedge\tilde e^\alpha=0\,,\qquad\tilde{\cal D}\tilde e^{\alpha}=0\,,\qquad\dot{\tilde e}^{\alpha}+K^{\alpha}=0\,,\label{lT.1}
\fequ
and Einstein's equations
\eqn
\tilde W_{\alpha}\wedge\tilde e^{\alpha}=0\,,\qquad\epsilon_{\alpha\beta\gamma}\tilde{\cal D}K^\beta\wedge\tilde e^\gamma=0\,,\qquad\tilde W_{\alpha}+\epsilon_{\alpha\beta\gamma}\left(\dot K^\beta+\frac1{\ell^2}\tilde e^\beta\right)\wedge\tilde e^\gamma=0\,.\label{lE}
\feqn
An important role is played by the quantity $\tilde W_{\alpha}$ defined in \eqref{tildeW} which is a component of the on-shell Weyl tensor\footnote{Details are given in \ref{app.Weyl}.}
\begin{eqnarray*}
W^{ab}=R^{ab}+\Lambda e^a\wedge e^b\,.
\end{eqnarray*}
Within our formalism and gauge-fixing the on-shell Weyl  tensor reads
\eqn
\sigma_{\perp}W^{0\alpha}&=& \dd t\wedge\left(\dot K^\alpha+\frac1{\ell^2}\tilde e^\alpha\right)+\tilde{\cal D}K^\alpha\,,\label{l.Weyl.1}\\
W^\alpha&=&\frac{\sigma_\perp}2\epsilon^\alpha{}_{\beta\gamma}W^{\beta\gamma}=\dd t\wedge \dot B^\alpha+\tilde W^\alpha\,.\label{l.Weyl.2}
\feqn

\section{The initial value formulation of gravity}
\label{initval}
The 3+1 split formalism can be nicely used towards the initial value formulation of ge\-ne\-ral re\-la\-ti\-vi\-ty which deals with the definition of a well-posed Cauchy problem for Einstein equations. Here we refer to the standard notion of an initial value problem, describing the time evolution of a set of initial data on a Cauchy spacelike surface, only for positive or vanishing cosmological constant. In the case of a negative cosmological constant the 3+1 split formalism describes instead the evolution of certain data on a Lorentzian hypersurface along a spacelike transverse (radial) coordinate. We gather this two physically distinct problems into the same conceptual framework of the ``initial value problem'' since the mathematics endowed in the 3+1 split formalism makes us act this way naturally. Firstly, the setup is the right one: we deal with a four dimensional ma\-ni\-fold sliced by three dimensional submanifolds $\Sigma_t$, parameterized by the coordinate $t$, which are naturally endowed with a metric structure defined by a vielbein $\tilde e^\alpha$ and torsionless spin connection $B^\alpha$ whose curvature $\rho^\alpha$ is the Riemann on the slice. Picking up a particular $t_0$, the submanifold $(\Sigma_{t_0},\tilde e^\alpha_{t_0},B^\alpha_{t_0})$ can suitably play the role of ``initial position'' in the Cauchy problem. 

Moreover, an additional symmetric\footnote{Symmetry and, later on, trace properties of various one-forms refer to their components e.g. for the vector valued one-form $V^\alpha =V^\alpha{}_{\beta}\tilde{e}^\beta$ the symmetry and trace properties refer to $V^\alpha{}_{\beta}$.} (see the first equation of (\ref{lT.1}))   field $K^\alpha$ exists and, by virtue of the third equation of \eqref{lT.1}, it is related to the velocity of the immersion of the vielbein towards the transverse $t$-direction. Hence its value on the slice $\Sigma_{t_0}$ must play the role of ``initial velocity'', or exstrinsic curvature in geometrical terms.

The remaining equations are the ``{\it dynamical equation}'' -- the third of \eqref{lE} -- which involves the ``acceleration'', i.~e.~the derivative of the $K^\alpha$, and the ``{\it integrability conditions}'' -- the first two equations of \eqref{lE} -- which are algebraical in the sense that the coordinate $t$ appears as a parameter and hence they are valid on any slice.

Using our definitions we can thus reformulate the theorem 10.2.2 in \cite{Wald:GR}, chapter~10, page~264, in the presence of a cosmological constant:
\Theo{Initial Value Formulation}{Consider a three-dimensional smooth manifold $\Sigma$ with signature (3,0) (when $\sigma_\perp=-1$) or (2,1) (when $\sigma_\perp=1$),  together with a metric structure defined by a vielbein $\varepsilon^\alpha$ and its torsionless spin connection $b^\alpha$, say $\tilde{\cal D}_b\varepsilon^\alpha=0$, and a 1-form $\kappa^\alpha$ satisfying a symmetry constraint $\kappa_\alpha\wedge\varepsilon^\alpha=0$. If the metric structure $(\varepsilon^\alpha,b^\alpha)$ and the additional field $\kappa^\alpha$ satisfy the following conditions
\equ
\tilde w_{\alpha}\wedge\varepsilon^{\alpha}=0\,,\qquad\epsilon_{\alpha\beta\gamma}\tilde{\cal D}_b \kappa^\beta\wedge\varepsilon^\gamma=0\,,\label{int.theo}
\fequ
where
\begin{eqnarray*}
\tilde w_\alpha=\tilde\dd b_{\alpha}+\frac12\epsilon_{\alpha\beta\gamma}b^\beta\wedge  b^\gamma-\frac12\epsilon_{\alpha\beta\gamma}\kappa^\beta\wedge \kappa^\gamma+\frac1{2\ell^{2}}\epsilon_{\alpha\beta\gamma}\varepsilon^\beta\wedge \varepsilon^\gamma\,,
\end{eqnarray*}
then there exists a \emph{unique} four-dimensional spacetime $(\M,g)$ of signature $(3,1)$ 
satisfying the following properties:
\begin{enumerate}
\item The metric $g$ is given by
\begin{eqnarray*}
g=\sigma_\perp\dd t\otimes\dd t+\tilde e^\alpha\otimes\tilde e_\alpha\,,
\end{eqnarray*}
where $\lim_{t\to t_0}\tilde e^\alpha=\varepsilon^\alpha$. Moreover at any slice we define the torsionless spin connection $B^\alpha$ of $\tilde e^\alpha$, with $\lim_{t\to t_0}B^\alpha=b^\alpha$. The extrinsic curvature $K^\alpha$ of the foliation also satisfies $\lim_{t\to t_0}K^\alpha=k^\alpha$.
\item $(\M,g)$ satisfies Einstein's equations with cosmological constant \begin{eqnarray*}\Lambda_{\rm cosm.}=-\frac{3\sigma_\perp}{\ell^2}\,,\end{eqnarray*} which means that $\{\tilde e^\alpha,\,B^\alpha,\,K^\alpha\}$ satisfy the integrability conditions \eqref{int.theo} on any slice $\Sigma_t$, (the first two of \eqref{lE}), and the dynamical equation (the third of \eqref{lE}).
\item Every other spacetime $(\M',g')$ satisfying (i)--(iii) can be mapped isometrically into a subset of $(\M,g)$. Furthermore $(\M,g)$ satisfies the domain of dependence property (as explained in the chapter of the textbook we referred to before, but we do not enter in the details here).
\end{enumerate}
}
Note that the standard initial value formulation corresponds to the limit for vanishing cosmological constant. In this case $t$ is the real time and $\Sigma=\Sigma_{t_0}$ is a Cauchy surface; moreover the spacetime $({\cal M},g)$ is globally hyperbolic. In the other cases the global hyperbolicity ceases to be a necessary condition; for instance if $\sigma_\perp=1$ global hyperbolicity of the four-dimensional spacetime is lost (see \cite{IshiWald1,IshiWald2} for a discussion of the simple AdS example). 

The idea is to extend this kind of description to the boundary. This is a particular slice, $\partial\M=\Sigma_\infty$, which is reached when the transverse coordinate $t$ takes its boundary value (typically $t=\pm\infty$). In other words, the boundary is the slice that can $t$-evolve {\it only} backwards (forward). Any bulk solution induces a three-dimensional metric on the slices $\Sigma_t$ for every $t$. However, given a bulk solution only a conformal class of metrics can be specified at the boundary \cite{FG,WittenAdSCFT}. One can then pick a particular representative boundary metric by choosing a defining function. Hence the correct ``initial position'', as we are putting it, is given by a certain conformal class. Different bulk geometries  arise by $t$-evolution by giving some ``initial velocity'' to the initial conformal data. Nevertheless, the ``initial velocity" need not be conformally invariant. 

Since our initial value problem is formulated at the boundary it should somehow be related to holography. Indeed, in the next subsection we will show that the different methods of holographic renormalization correspond to different ways of setting up an initial value problem at the boundary i.e. different ways to define the appropriate ``initial boundary velocity".  

\section{The Fefferman-Graham expansion in the 3+1 split formalism}
\label{FG.holo}
The Fefferman-Graham (FG) expansion of the metric \cite{FG,fefferman-2007} has proven to be the most efficient method in holographic applications \cite{dHSS}. Here we  present a detailed transcription of the FG expansion to all our 3+1-split quantities. In doing so, we discover that the coefficients in the FG expansion are intimately related to the geometrical data of the boundary. 

In the FG expansion the vielbein is expanded in powers of $e^{-t/\ell}$ as
\equ
\label{FGvielbein}
\tilde e^\alpha=e^{t/\ell}E^\alpha(x)+e^{-t/\ell}\sum_{k=0} F^\alpha_{[k+2]}(x)e^{-kt/\ell}\,.\label{FG.vielbein}
\fequ
In the absence of sources the finite term in the expansion (\ref{FGvielbein}) above is missing \cite{Skenderis:2002wp}, hence we neglect it right from the beginning. We will be interested in the boundary at $t=+\infty$ where  $E^\alpha$ is a representative of the conformal class of boundary vielbeins. Recall that $t$ is related to the standard Poincar\'e patch radial coordinate $r\in [0,\infty)$ as $r/\ell=e^{-t/\ell}$. Picking then a particular defining function we can refer to $E^\alpha$ as the {\it boundary vielbein}. 

The electric and the magnetic fields are obtained by solving the equations \eqref{lT.1}. From the  third equation in \eqref{lT.1} we find 
\equ
K^{\alpha}=-\frac1\ell e^{t/\ell}E^{\alpha}+\frac1\ell e^{-t/\ell}\sum_{k=0}(k+1)F^{\alpha}_{[k+2]}e^{-kt/\ell}\,.\label{FG.electric}
\fequ
The first equation of \eqref{lT.1} determines the symmetry properties of the components of the FG expansion. The first few orders yield
\equ
F^{\alpha}_{[2]}\wedge E_{\alpha}=F^{\alpha}_{[3]}\wedge E_{\alpha}=F^{\alpha}_{[4]}\wedge E_{\alpha}=0\,.\label{FG.symm}
\fequ
The magnetic field has the expansion
\begin{eqnarray*}
B^{\alpha}=\sum_{k=0}B^{\alpha}_{[k]}e^{-kt/\ell}\,,
\end{eqnarray*}
and has a finite  $t\rightarrow\infty$ limit. Its various coefficients  in the expansion are implicitly obtained by solving the second equation of \eqref{lT.1} and the first few orders give
\eqn
{\cal D}_{[0]}E^{\alpha}=B^\alpha_{[1]}=0\,,\label{mag.FG.1}\\
{\cal D}_{[0]}F^{\alpha}_{[2]}+\epsilon^{\alpha}{}_{\beta\gamma}B^{\beta}_{[2]}\wedge E^{\gamma}=0\,,\label{mag.FG.3}\\
{\cal D}_{[0]}F^{\alpha}_{[3]}+\epsilon^{\alpha}{}_{\beta\gamma}B^{\beta}_{[3]}\wedge E^{\gamma}=0\,,\label{mag.FG.4}\\
{\cal D}_{[0]}F^{\alpha}_{[4]}+\epsilon^{\alpha}{}_{\beta\gamma}B^{\beta}_{[2]}\wedge F^{\gamma}_{[2]}+\epsilon^{\alpha}{}_{\beta\gamma}B^{\beta}_{[4]}\wedge E^{\gamma}=0\,,\label{mag.FG.5}
\feqn
where ${\cal D}_{[0]}$ is the three-dimensional covariant exterior derivative made with the leading order magnetic field $B^{\alpha}_{[0]}$. From \eqref{mag.FG.1} we learn that $B^{\alpha}_{[0]}$ is the torsionless spin connection of the boundary vielbein $E^\alpha$. By taking the exterior multiplication of \eqref{mag.FG.3}--\eqref{mag.FG.4} with $E_\alpha$ we find
\begin{eqnarray*}
\epsilon_{\alpha\beta\gamma}B_{[2]}^\alpha\wedge E^\beta\wedge E^\gamma=0\,,\qquad\epsilon_{\alpha\beta\gamma}B_{[3]}^\alpha\wedge E^\beta\wedge E^\gamma=0\,,
\end{eqnarray*}
and hence the  components $B^\alpha_{[2]}$ and $B^\alpha_{[3]}$ of the magentic field are traceless.

Next we need to solve Einstein's equations, order by order in the FG expansion. To do that we first compute the components of the Weyl tensor as
\eqn
\dot K^{\alpha}+\frac1{\ell^{2}}\tilde e^{\alpha}&=&-\frac1{\ell^2}\sum_{k=0}^\infty[(k+2)^2-1]F^\alpha_{[k+3]}e^{-(k+2)t/\ell}\,,\label{FG.Weyl.expansion.1}\\
\tilde{\cal D}K^{\alpha}&=&-e^{-t/\ell}\frac2\ell\epsilon^{\alpha}{}_{\beta\gamma}B^{\beta}_{[2]}\wedge E^{\gamma}-e^{-2t/\ell}\frac3\ell \epsilon^{\alpha}{}_{\beta\gamma}B^{\beta}_{[3]}\wedge E^{\gamma}+{\cal O}\left(e^{-3t/\ell}\right)\,,\label{FG.Weyl.expansion.2}\\
\dot B^\alpha&=&-\frac1\ell\sum_{k=0}^\infty(k+2)B^{\alpha}_{[k+2]}e^{-(k+2)t/\ell}\,,\label{FG.Weyl.expansion.3}\\
\tilde W^{\alpha}&=&\rho^{\alpha}_{[0]}+\frac2{\ell^{2}}\epsilon^{\alpha}{}_{\beta\gamma}F^{\beta}_{[2]}\wedge E^{\gamma}+e^{-t/\ell}\frac3{\ell^{2}}\epsilon^{\alpha}{}_{\beta\gamma}F^{\beta}_{[3]}\wedge E^{\gamma}\nonumber\\
&&\qquad\qquad+e^{-2t/\ell}\left[{\cal D}_{[0]}B^\alpha_{[2]}+\frac4{\ell^2}\epsilon^\alpha{}_{\beta\gamma}F^\beta_{[4]}\wedge E^\gamma\right]+{\cal O}\left(e^{-3t/\ell}\right)\,.\label{FG.Weyl.expansion.4}
\feqn
Hence, Einstein's equations yield:
\begin{description}
\item[{\bf Equation $\epsilon_{\alpha\beta\gamma}\tilde{\cal D}K^{\beta}\wedge\tilde e^{\gamma}=0$:}] This is the equivalent of  Gauss law and to leading and subleading order simply imposes the symmetry properties of the components of the magnetic field
\equ
B^{\alpha}_{[2]}\wedge E_{\alpha}=0\,,\qquad B^{\alpha}_{[3]}\wedge E_{\alpha}=0\,.\label{FG.eom.1}
\fequ

\item[{\bf Equation $\tilde W_{\alpha}+\epsilon_{\alpha\beta\gamma}\left(\dot K^{\beta}+\frac1{\ell^{2}}\tilde e^{\beta}\right)\wedge\tilde e^{\gamma}=0$:}] This is what we have called the {\it dynamical equation}. It reveals that the various coefficients in the FG expansion correspond to geometrical quantities of the boundary.  The equation reads 
\eqn
\rho_{[0]}^{\alpha}+\frac2{\ell^{2}}\epsilon^\alpha{}_{\beta\gamma}F^{\beta}_{[2]}\wedge E^{\gamma}+e^{-2t/\ell}\left[{\cal D}_{[0]}B^\alpha_{[2]}-\frac4{\ell^2}\epsilon^\alpha{}_{\beta\gamma}F^\beta_{[4]}\wedge E^\gamma\right]+{\cal O}\left(e^{-3t/\ell}\right)=0\,.\label{FG.eom.2}
\feqn
To leading order it gives
\equ
\rho_{[0]}^{\alpha}+\frac2{\ell^{2}}\epsilon^\alpha_{\,\,\beta\gamma}F^{\beta}_{[2]}\wedge E^{\gamma}=0\,,\label{l.Schouten}
\fequ
and hence it shows that $F^\alpha_{[2]}$ is proportional to the boundary Schouten tensor, whose details are given in \ref{app.Weyl}. This follows from the fact that the three-dimensional equation $\Lambda_\alpha+\epsilon_{\alpha\beta\gamma}F^\beta\wedge E^\gamma=0$ can be solved for the one-form  $F^\alpha$ in terms of the  Hodge dual of the two-form $\Lambda_\alpha$, provided $E^\alpha$ is a vielbein and hence invertible. Explicitly, if $F^\alpha=F^\alpha{}_\beta E^\beta$, we have that $F^\alpha{}_\beta-{\rm tr}(F)\delta^\alpha{}_\beta=-\sigma_\perp({}^{\tilde*}\Lambda_\beta)^\alpha$. Moreover, from \eqref{l.Schouten} the tensor $F_{[2]}-{\rm tr}(F_{[2]}){\rm id.}$ is proportional to the Einstein tensor, as it should be. Explicitly we have
\begin{eqnarray*}
-\frac{2\sigma_{\perp}}{\ell^{2}}F^{\alpha}_{[2]}={}^{(3)}S^{\alpha}={\rm Ric}^{\alpha}-\frac{R}4E^{\alpha}\,,
\end{eqnarray*}
where ${\rm Ric}^{\alpha}=E_{\beta}\rfloor\rho^{\beta\alpha}$ and $R=E_{\alpha}\rfloor{\rm Ric}^{\alpha}$. For the same reason, from \eqref{mag.FG.3} $B^{\alpha}_{[2]}$ is given in terms of the Hodge dual of the boundary Cotton-York tensor. Since  $B^\alpha_{[2]}$ is symmetric and traceless we obtain
\begin{eqnarray*}
B^\alpha_{[2]}=-\sigma_\perp{}^{\tilde*}{\cal D}_{[0]}F^\alpha_{[2]}=\frac{\ell^{2}}2{}^{\tilde*}C^{\alpha}\,,
\end{eqnarray*}
where $C^{\alpha}={\cal D}_{[0]}{}^{(3)}S^{\alpha}$. In three dimensions the Cotton-York tensor is the only irreducible conformally invariant tensor \cite{graham-2004}. It vanishes if and only if the metric is conformally flat.  Since $F^\alpha_{[2]}$ and $B^\alpha_{[2]}$ are related to each other, the $e^{-2t/\ell}$-component of \eqref{FG.eom.2} relates $B^\alpha_{[2]}$ to $F^\alpha_{[4]}$ as
\equ
F^\alpha_{[4]}-{\rm tr}(F_{[4]})E^\alpha=\sigma_\perp\frac{\ell^2}4{}^{\tilde*}{\cal D}_{[0]}B^\alpha_{[2]}=\sigma_\perp\frac{\ell^4}8{}^{\tilde*}{\cal D}_{[0]}{}^{\tilde*}C^{\alpha}\,.\label{FG.Bach.l}
\fequ
Moreover, the three-dimensional spatial component of the Weyl tensor reads
\eqn
\tilde W^\alpha=e^{-t/\ell}\frac3{\ell^{2}}\epsilon^{\alpha}{}_{\beta\gamma}F^{\beta}_{[3]}\wedge E^{\gamma}\nonumber\\
\qquad\qquad+e^{-2t/\ell}\frac8{\ell^2}\epsilon^\alpha{}_{\beta\gamma}F^\beta_{[4]}\wedge E^\gamma+{\cal O}\left(e^{-3t/\ell}\right)\,,\label{FG.Weyl.3D}
\feqn
and thus vanishes at the boundary.

\item[{\bf Equation $\tilde W_{\alpha}\wedge\tilde e^{\alpha}=0$:}] This is an algebraic equation whose leading and subleading terms set to zero the traces of the matrices $F^{\alpha}_{[3]\;\beta}$ and $F^{\alpha}_{[4]\;\beta}$, since it gives
\equ
\frac3{\ell^{2}}\epsilon_{\alpha\beta\gamma}F^{\alpha}_{[3]}\wedge E^{\beta}\wedge E^{\gamma}+e^{-t/\ell}\frac8{\ell^{2}}\epsilon_{\alpha\beta\gamma}F^{\alpha}_{[4]}\wedge E^{\beta}\wedge E^{\gamma}+{\cal O}\left(e^{-2t/\ell}\right)=0\,.\label{FG.eom.34}
\fequ
As a result \eqref{FG.Bach.l} is modified to
\begin{eqnarray*}
F^\alpha_{[4]}=\sigma_\perp\frac{\ell^4}8{}^{\tilde*}{\cal D}_{[0]}{}^{\tilde*}C^{\alpha}\,.
\end{eqnarray*}
\end{description}
A nice consequence of the above results  is that the whole Weyl tensor vanishes at the boundary,
\begin{eqnarray*}
\left.W^{ab}\right|_{\partial\M}=0\,.
\end{eqnarray*}
From the relationship between the on-shell Weyl tensor and the curvature one can easily find the asymptotic behavior of the latter e.g.  in order to define asymptotic charges.

Not all coefficients of the FG expansion are determined in terms of the boundary geometrical data $E^\alpha$ and $B^\alpha_{[0]}$. The quantity $F^{\alpha}_{[3]}$ is an independent coefficient. Actually, it is only possible to say that it is symmetric
\begin{eqnarray*}
F^{\alpha}_{[3]}\wedge E_{\alpha}=0\,,
\end{eqnarray*}
traceless
\begin{eqnarray*}
\epsilon_{\alpha\beta\gamma}F^{\alpha}_{[3]}\wedge E^{\beta}\wedge E^{\gamma}=0\,,
\end{eqnarray*}
and that it obeys a conservation law
\begin{eqnarray*}
\epsilon_{\alpha\beta\gamma}{\cal D}_{[0]}F^{\beta}_{[3]}\wedge E^\gamma=0\,.
\end{eqnarray*}
In a holographic setup this function determines the vacuum expectation value of the energy momentum tensor in the boundary conformal field theory. 

The fact that the general solution to Einstein's equations requires two different sets of undetermined data, $\{E^\alpha,F^\alpha_{[3]}\}$, is clearly related to the initial value problem of general relativity. In particular, we can naturally associate $E^\alpha$ to the {\it boundary initial position} and $F^\alpha_{[3]}$ to the {\it  boundary initial velocity} of the well-posed Cauchy problem that describes the transverse ``propagation'' of the boundary geometrical data towards the 4D bulk. In that sense, the vev of the energy momentum tensor of the boundary CFT can be viewed as an initial velocity.  

\section{Renormalization methods vs transformations of the canonical variables}
Having in mind the holographic application of our results we focus henceforth in the case where $\sigma_\perp=1$, namely the case where our bulk configurations are asymptotically Anti-de Sitter. However, we will continue using $\sigma_\perp$, in order to keep  track of the signature dependance of our results which can be used   in applications, other than holography, involving asymptotically de-Sitter spacetimes. 

In the  3+1 split formalism described in section \ref{initval}, it appears that the two canonically conjugate fields that describe ``position" and ``velocity" are given by $\{\tilde e^{\alpha},K^{\alpha}\}$. This is correct on any slice $\Sigma_{t}$ other than the boundary i.e. with $t\neq \infty$, where these quantities are finite. However, to define the correct geometrical data on the boundary one needs to multiply $\tilde{e}^\alpha$ and $K^\alpha$ by $e^{-t/\ell}$ and then take the $t\rightarrow\infty $ limit \cite{WittenAdSCFT}.  In this case, from \eqref{FG.vielbein} and \eqref{FG.electric} one finds that 
\equ
\label{KErelation}
K^{\alpha}=-\frac1\ell\tilde e^{\alpha}+{\cal O}\left(e^{-t/\ell}\right)\,
\fequ
hence it would seem that the boundary geometrical data extracted form $\tilde{e}^\alpha$ and $K^\alpha$ are proportional to each other. In fact, both the vielbein $\tilde e^\alpha$ and the extrinsic curvature $K^\alpha$ could suitably play the role of the boundary initial position.  The question is what plays the role of the {\it boundary initial velocity}. Looking at the expansions of the fields given in the previous section we note that the three-dimensional component of the on-shell Weyl tensor $\tilde W^{\alpha}$ has the expansion 
\equ
\label{tildeWexp}
\tilde W^\alpha=e^{-t/\ell}\frac3{\ell^{2}}\epsilon^{\alpha}{}_{\beta\gamma}F^{\beta}_{[3]}\wedge E^{\gamma}+{\cal O}\left(e^{-2t/\ell}\right)\,.
\fequ
Defining the one-form ${\cal P}^{\alpha}$ as
\begin{eqnarray*}
\tilde W^{\alpha}=\sigma_{\perp}\epsilon^{\alpha}{}_{\beta\gamma}{\cal P}^{\beta}\wedge\tilde e^{\gamma}
\end{eqnarray*}
its leading behavior is given by  $F^{\alpha}_{[3]}$ and hence it could nicely play the role of boundary initial velocity. 
To see that note that on any slice $\Sigma_{t}$, ${\cal P}_\alpha$ is symmetric ${\cal P}_{\alpha}\wedge\tilde e^{\alpha}=0$, due to the Bianchi identity, and traceless ${}^{\tilde*}{\cal P}_{\alpha}\wedge\tilde e^{\alpha}=0$, by virtue of the first of \eqref{lE}. In fact we have  ${\cal P}^{\alpha}={}^{\tilde*}\tilde W^{\alpha}$ and hence
\begin{eqnarray*}
{\cal P}^\alpha =\sigma_\perp\frac{3}{\ell^2}e^{-2t/\ell}F^\alpha_{[3]}+O\left(e^{-3t/\ell}\right)
\end{eqnarray*}
Also notice that ${\cal P}^\alpha$ is not in general conserved for $t\neq\infty$, but it becomes conserved at the boundary due to
\begin{eqnarray*}
\lim_{t\to+\infty}e^{t/\ell}\epsilon_{\alpha\beta\gamma}\tilde{\cal D}{\cal P}^{\beta}\wedge\tilde e^{\gamma}=0\,.
\end{eqnarray*}

The discussion above implies that both pairs of conjugate variables 
\equ
\{\tilde e^{\alpha},{\cal P}^{\alpha}\}\,,\qquad\{K^{\alpha},{\cal P}^{\alpha}\}\,,\label{truecanonical}
\fequ
can equivalently describe an initial value formulation at the three-dimensional boundary.  It appears therefore that the boundary is the point where holographic renormalization me\-thods meet the initial value formulation. Namely, starting from a Dirichlet problem where $\delta\tilde{e}^\alpha=0$ at the boundary, i.e. where $\tilde{e}^\alpha$ is the boundary initial position, one needs to make sure that the boundary initial velocity is ${\cal P}^\alpha$. This can be achieved by a transformation of the canonical momentum such that $K^\alpha\mapsto {\cal P}^\alpha$. We will show below that this procedure coincides with the standard holographic renormalization (i.e. \cite{BK,dHSS}). On the other hand, one could have started with a Dirichlet problem where $\delta K^\alpha =0$ at the boundary i.e.  $K^\alpha$ being the  boundary initial position. This is equivalent to {\it not} adding the Gibbons-Hawking term in the Einstein-Hilbert action. Again, one needs to make sure that the boundary initial velocity is given by ${\cal P}^\alpha$ and this can be achieved by the transformation $\tilde{e}^\alpha\mapsto {\cal P}^\alpha$, or equivalently $\Sigma_\alpha \mapsto \tilde{W}_\alpha$. We will demonstrate below that this second procedure coincides with the method of Kounterterms \cite{Zanelli,Mora:2004kb,Olea} where the infinities are cancelled by the addition of the Euler density.

To be explicit, the essence of holography is the evaluation of the on-shell gravitational action which is then identified with (minus) the generating functional of connected diagrams of a boundary  conformal field theory in the leading saddle point approximation. The boundary values of bulk fields are interpreted as external sources for  boundary conserved currents. For pure gravity in the bulk, we have schematically
\equ
\label{Holrecipe}
S\Bigl|_{{\rm os}}[E^{\alpha}]=-W_{\rm QFT}[E^{\alpha}]\,.
\fequ
Since $E^\alpha$ plays the role of an external source in the boundary, the variation $\delta S\Bigl|_{{\rm os}}$ must be zero for fixed $E^\alpha$.  This is equivalent to the statement of ensuring a well posed Dirichlet problem for the vielbein, hence a natural starting point for holography is the gravitational action $S=S_{{\rm EH}}+S_{{\rm GH}}$. Schematically, indicating $\{\lambda_{i}\}=\{N,N^{\alpha},q^{0\alpha},Q^{\alpha}\}$ the lagrange multipliers providing the constraints ${\cal C}^{i}$ respectively, the gravitational action reads 
\equ
\label{SEHGHos}
S=\frac{\sigma_\perp}{8\pi G}\int_\M \left[\epsilon_{\alpha\beta\gamma} K^\alpha\wedge \tilde{e}^\beta\wedge \dd\tilde{e}^\gamma-\lambda_{i}{\cal C}^{i}\right]\,.
\fequ
Its on-shell variation  reads 
\equ
\delta S\Bigl|_{{\rm os}}=\frac{\sigma_\perp}{8\pi G}\int_{\partial\M}\epsilon_{\alpha\beta\gamma}K^\alpha\wedge\tilde e^\beta\wedge\delta\tilde e^\gamma\,,\label{OS.var}
\fequ
and hence the presence of the Gibbons-Hawking term ensures that  only the variation with respect to the vielbein survives.

However, one could have {\it not} added the Gibbons-Hawking term, in which case we would consider simply the Einstein-Hilbert action \eqref{lor.act} which schematically reads
\equ\label{SEHos}
S_{{\rm EH}}=-\frac{\sigma_\perp}{8\pi G}\int_\M\left[\Sigma_\alpha\wedge \dd K^\alpha+\lambda_{i}{\cal C}^{i}\right]\,.
\fequ
Its on-shell variation is given by
\equ
\delta S_{{\rm EH}}\Bigl|_{{\rm os}}=-\frac{\sigma_\perp}{8\pi G}\int_{\partial\M}\Sigma_\alpha\wedge \delta K^\alpha\,.\label{OS.var1}
\fequ
Since $\tilde{e}^\alpha$ and $K^\alpha$ are proportional to each other at the boundary, both the starting points  (\ref{OS.var}) and (\ref{OS.var1}) correspond to the {\it same} Dirichlet problem and hence are expected to correspond to the same boundary physics. Moreover, in both cases we need to ensure that the initial boundary velocity is the same, given by the boundary value of ${\cal P}^\alpha$, and here lies the difference between the two cases; we need different transformations to achieve that. We will show below that the two different transformations leading to the same boundary initial velocity correspond to the standard holographic renormalization \cite{BK,dHSS} and to the Kounterterms method  \cite{Zanelli,Mora:2004kb,Olea} respectively.  

\subsection{Holographic renormalization}
The problem that we need to take care of in (\ref{OS.var}) can be phrased in two different ways; we can {\it either}  say that the two-from $\epsilon_{\alpha\beta\gamma}K^\beta\wedge\tilde{e}^\gamma $ is not a well-defined initial velocity {\it or} we can say that (\ref{OS.var}) diverges at the boundary. These two points of view are essentially equivalent. Indeed, by virtue of the definition (\ref{tildeW}) and the expansions (\ref{KErelation}), (\ref{tildeWexp}) we notice that the quantity  $\tilde{W}'_\alpha$ defined as
\equ
\label{holrenorm}
\epsilon_{\alpha\beta\gamma}K^\beta\wedge\tilde{e}^\gamma \equiv\ell \tilde{W}'_\alpha-\ell\rho_\alpha-\frac{2}{\ell}\Sigma_\alpha\,,
\fequ
has the same near boundary expansion as $\tilde{W}_\alpha$, namely
\equ
\label{tildeWprime}
\tilde{W}'_\alpha =e^{-t/\ell}\frac{3}{\ell^2}\epsilon_{\alpha\beta\gamma}F^\beta_{[3]}\wedge E^\gamma +O\left(e^{-2t/\ell}\right)
\,.
\fequ
Hence, our strategy is to implement the transformation (\ref{holrenorm}) at the level of the action. Then we will be sure that the new canonical momentum $\tilde{W}'_\alpha$ will give on shell the proper boundary initial velocity. 

We implement the transformation (\ref{holrenorm}) on the restricted phase space defined by the constraints i.e. the fields appearing in (\ref{holrenorm}) satisfy the constraints. Then, the insertion of (\ref{holrenorm}) into (\ref{SEHGHos}) modifies the gravitational action, when we set to zero the constraints, as
\equ
\label{S'os}
S=\frac{\sigma_\perp\ell}{8\pi G}\int_{\cal M} \tilde{W}'_\alpha\wedge \dd\tilde{e}^\alpha-\frac{\sigma_\perp\ell}{8\pi G}\int_{\partial\M}\left[\dot B_{\alpha}\wedge\tilde e^{\alpha}\wedge\dd t+\rho_\alpha\wedge{\tilde e}^\alpha+\frac{1}{3\ell^2}\epsilon_{\alpha\beta\gamma}{\tilde e}^\alpha\wedge{\tilde e}^\beta\wedge{\tilde e}^\gamma\right]\,.
\fequ
The first term inside the brackets of (\ref{S'os}) vanishes by virtue of the third of \eqref{lT.1} together with the second of \eqref{lE}. Hence, the transformation (\ref{holrenorm}), when implemented on the restricted phase space defined by the constraints, modifies the  action by boundary terms. In fact, one should be able to show that the constraints are not modified and hence that (\ref{holrenorm}) is a proper canonical transformation.  The two boundary terms  are geometrical quantities, namely the curvature and the volume form defined on the slice. They coincide with {\it minus} the original counterterms used in the context of holographic renormalization \cite{BK,dHSS}. We can subtract them to be left with the so-called renormalized action $S'_{ren}$ whose on-shell variation yields at the boundary  
\equ
\label{dSholrenorm}
\delta S'_{\rm ren.}\Bigl|_{{\rm os}}=\frac{3\sigma_\perp}{8\pi G\ell}\int_{\partial\M}\epsilon_{\alpha\beta\gamma}F^\alpha_{[3]}\wedge E^\beta\wedge\delta E^\gamma+{\cal O}(e^{-t/\ell})\,.
\fequ
The holographic interpretation of (\ref{dSholrenorm}) is that the expectation value of the boundary energy momentum tensor is  related to the Hodge dual of the Weyl tensor as
\begin{eqnarray*}
\tau_\alpha\equiv\frac{\delta S'_{{\rm ren.}}}{\delta E^\alpha}=\frac{3\sigma_\perp}{8\pi G\ell}\epsilon_{\alpha\beta\gamma}F^\beta_{[3]}\wedge E^\gamma=\frac{\sigma_\perp\ell}{8\pi G} \lim_{t\to+\infty}e^{t/\ell}\tilde W_{\alpha}\,,
\end{eqnarray*}
and hence explicitly 
\equ
\langle T_{ij}\rangle_{s}=E^\alpha{}_i\left({}^{\tilde*}\tau_\alpha\right)_j=\frac3{8\pi G\ell}F_{[3]\;ij}\,.\label{vev.stress.general}
\fequ
Recall that  $F_{[3]\;\alpha\beta}$ is traceless, symmetric and conserved, as the energy momentum tensor of a three-dimensional CFT should be. 

\subsection{Kounterterms}
We can now try to set up an initial value formalism for gravity starting with the  Einstein-Hilbert action, without adding to it the Gibbons-Hawking term. In this case we wold need to make the transformations   
\equ
\Sigma_{\alpha}=\ell^{2}\tilde W_{\alpha}-\ell^{2}\rho_{\alpha}+\frac{\ell^{2}}2\epsilon_{\alpha\beta\gamma}K^{\beta}\wedge K^{\gamma}\,,
\fequ
into (\ref{SEHos}), setting to zero the constraints. We get
\eqn
\label{kounter}
S_{{\rm EH}}\mapsto S'_{{\rm EH}}&=&-\frac{\sigma_\perp\ell^{2}}{8\pi G}\int_{\cal M}\left[\tilde{W}_\alpha\wedge \dd K^\alpha+\tilde{\cal D}K_{\alpha}\wedge\dd B^{\alpha}\right]\nonumber\\
&&+\frac{\sigma_\perp\ell^{2}}{8\pi G}\int_{\partial\M}\left[\rho_\alpha\wedge K^\alpha-\frac16\epsilon_{\alpha\beta\gamma}K^\alpha\wedge K^\beta\wedge K^\gamma\right]\,.\nonumber\\
\feqn
where, in the boundary integral, we have already dropped a term proportional to
\begin{eqnarray*}
\int_{\partial\M}\dot B_{\alpha}\wedge K^{\alpha}\wedge\dd t
\end{eqnarray*}
which vanishes since the leading term of the integrand is proportional to $e^{-t/\ell}$. The second  term in the first line of \eqref{kounter},  gives the on shell contribution  $\int_{\partial\M}\tilde{\cal D}K_{\alpha}\wedge\delta B^{\alpha}$, which vanishes identically at the $t=\infty$ boundary since the integrand is proportional to $e^{-t/\ell}$ (notice that the magnetic field is finite in the boundary). Dropping this term we are left with the two boundary contributions in the second line of (\ref{kounter}). Again, one should be able to show that the constraints do not change and that the transformation above is canonical.

Remarkably, the boundary term we are left will in (\ref{kounter})  are exactly   {\it minus} the Euler density
\eqn
\chi&=&-\frac{\sigma_{\perp}\ell^{2}}{64\pi G}\int\epsilon_{abcd}R^{ab}\wedge R^{cd}\nonumber\\
\phantom{\chi}&=&-\frac{\sigma_{\perp}\ell^{2}}{8\pi G}\int_{\partial\M}\left[\rho_{\alpha}\wedge K^{\alpha}-\frac16\epsilon_{\alpha\beta\gamma}K^{\alpha}\wedge K^{\beta}\wedge K^{\gamma}\right]\,,\label{Euler.term.l}
\feqn
hence adding it to \eqref{kounter} we would obtain the on-shell action 
\equ
S_{{\rm ren.}}\Bigl|_{os}=S'_{{\rm EH}}+\chi=-\frac{\sigma_\perp\ell^{2}}{8\pi G}\int_{\cal M}\tilde{W}_\alpha\wedge \dd K^\alpha\,,\label{eff.on.shell}
\fequ
As shown above, the variation of  \eqref{eff.on.shell} gives exactly the previous result \eqref{dSholrenorm} and hence the stress tensor is the same as in \eqref{vev.stress.general}.

We conclude that the two procedures, holographic renormalization and Kounterterms, can be equivalently used to setup an initial value formulation for gravity in the $t=\infty$ boundary and - as we propose - can be used equivalently for its holographic description. At this point we also notice that the Kounterterm method is intriguingly connected with the geometrical Mac-Dowell Mansouri formulation of gravity \cite{MacDowell:1977jt,Anabalon:2006fj,Anabalon:2007dr,Anabalon:2008hi,Wise:2006sm}. Indeed, the sum of the Einstein-Hilbert action plus the Euler density with the {\it exact} coefficient given in (\ref{Euler.term.l}) is the MM action 
\equ
S_{{\rm MM}}=-\frac{\sigma_{\perp}\ell^{2}}{64\pi G}\int\epsilon_{abcd}W^{ab}\wedge W^{cd}\,.\label{MMgravity}
\fequ
The two-form $W^{ab}$ coincides on-shell with the Weyl tensor which, as discussed at the end of  \ref{app.Weyl}, plays the role of the Lorentz component of the curvature of a ${\rm so}(3,2)$(${\rm so}(4,1)$)-valued connection for $\sigma_{\perp}=1$ ($\sigma_{\perp}=-1$). Hence, \eqref{MMgravity} coincides on-shell with the renormalized action (\ref{eff.on.shell}) and it also gives a procedure to compute {\rm finite} conserved quantities associated to spacetimes \cite{Zanelli}.

\section{Conclusions}
We presented a detailed analysis of gravity in the 3+1-split formalism having in mind applications to AdS$_4$/CFT$_3$ holography. The formalism allows for the setup of an initial value problem at the $t=\infty$ boundary. We presented the explicit Fefferman-Graham expansion of the various quantities involved and noted that their coefficients correspond to geometrical boundary data. Armed with our explicit results, we have argued that the holographic description of gravity can alternatively be  considered as the formulation of an initial value problem at the boundary. In this context we have shown that holographic renormalization and the Kounterterm method both correspond to certain transformations of the canonical variables. In the companion work \cite{MPT2} we will discuss the emergence of gravitational Chern-Simons in the boundary of four-dimensional gravity and also the consequences of self-duality in the case of Euclidean signature. 
We believe that our techniques and results can provide the basis for extensive studies in AdS$_4$/CFT$_3$ holography. 
Finally, our approach has many similarities with past work on quantum gravity\footnote{We thank Lee Smolin for bringing these works to our attention.}, in particular on its holographic formulation \cite{Smolin1,Smolin2,Smolin3}, and therefore it may be useful in linking the two fields. 

\section*{Acknowledgements}
We would like to thank D.~Klemm, G.~Kofinas, D.~Minic and R.~Olea for helpful discussions. A.~C.~P. would like to thank S.~de~Haro and R.~G.~Leigh for long term enlightening discussions and collaboration on similar ideas, and the latter for a critical reading of the manuscript. D.~M. and A.~C.~P. were partially supported by the INTERREG IIIA Program K2301.007. A.~C.~P. was partially supported by the European RTN Program MRTN-CT-2004-512194. G.~T. was partially supported by the Italian MIUR-PRIN contract.

\appendix

\section{Weyl's Conformal Tensor}
\label{app.Weyl}
Consider a four-dimensional manifold $\M$ endowed with  a metric structure described by a vielbein $e^a$ and a torsionless ${\rm so}(3,1)$-valued connection $\omega^{ab}$. The Riemann tensor, given explicitly as the curvature of the Lorentz connection, $R^{ab}=\dd\omega^{ab}+\omega^a{}_c\wedge\omega^{cb}$, can be decomposed into the following parts which are irreducible representations of the full Lorentz group
\equ
R^{ab}=C^{ab}+E^{ab}+G^{ab}\,,\label{Weyl.dec.def}
\fequ
where
\equ
E^{ab}=e^{[a}\wedge F^{b]}\,,\qquad G^{ab}=\frac{R}{12}e^a\wedge e^b\label{Weyl.dec.def2}
\fequ
being $F^a={\rm Ric}^a-\frac{R}4e^a$ the traceless part of the Ricci tensor $R^{a}{}_b$, ${\rm Ric}^a=R^a{}_be^b$ the Ricci 1-form and $R=e
_a\rfloor{\rm Ric}^a=R^a{}_a$ the scalar curvature. This decomposition defines the \emph{Weyl conformal tensor} $C^{ab}$: it is called ``conformal'' since its components do not change under conformal transformations. It is possible to define the Weyl tensor in any dimensions $D$ (actually for $D>3$) in the following way
\begin{eqnarray*}
C^{ab}\equiv R^{ab}-e^{a}\wedge S^{b}+e^b\wedge S^a\,,
\end{eqnarray*}
where
\begin{eqnarray*}
S^a\equiv\frac1{D-2}\left[{\rm Ric}^a-\frac{R}{2(D-1)}e^a\right]
\end{eqnarray*}
is the Schouten tensor. Besides the standard symmetries enjoyed by the Riemann tensor, the Weyl tensor has the additional feature to be completely traceless, $e_a\rfloor C^{ab}=0$, and hence in four dimensions it has ten independent components. In three dimensions it turns out that the Weyl tensor vanishes identically and thus the Riemann tensor is given entirely in terms of the Schouten tensor, 
\begin{eqnarray*}
{}^{(3)}R^{ab}=e^a\wedge {}^{(3)}S^b-e^b\wedge {}^{(3)}S^a\,,
\end{eqnarray*}
with
\begin{eqnarray*}
{}^{(3)}S^a={\rm Ric}^a-\frac{R}4e^a\,.
\end{eqnarray*}
Let us go back to the original definition \eqref{Weyl.dec.def}, given in the case of a four dimensional manifold. When we consider ${\rm so}(3,1)$-valued 2-forms $\Lambda^{ab}=\frac12\Lambda^{ab}{}_{cd}e^c\wedge e^d$, such as any term in \eqref{Weyl.dec.def}, we can deal with two different notions of Hodge duality: one concerning the flat, tangent indices
\equ
{}^{\hat*}\Lambda^{ab}=\frac12\epsilon^{ab}{}_{a'b'}\Lambda^{a'b'}=\frac14\epsilon^{ab}{}_{a'b'}\Lambda^{a'b'}{}_{cd}e^c\wedge e^d\,,\label{app.tang.hodge}
\fequ
and one concerning curved, spacetime indices
\begin{eqnarray*}
{}^*\Lambda^{ab}=\frac12\Lambda^{ab}{}_{c'd'}{}^*\left(e^{c'}\wedge e^{d'}\right)=\frac14\Lambda^{ab}{}_{c'd'}\epsilon^{c'd'}{}_{cd}e^c\wedge e^d\,.
\end{eqnarray*}
The two notions, in general, have nothing to do with each other. But, from the definitions we gave in \eqref{Weyl.dec.def2}, it turns out that
\begin{eqnarray*}
{}^{\hat*}C^{ab}={}^*C^{ab}\,,\qquad {}^{\hat*}E^{ab}=-{}^*E^{ab}\,,\qquad{}^{\hat*}G^{ab}={}^*G^{ab}\,.
\end{eqnarray*}
If Einstein's equations hold, in absence of any source term, ${\rm Ric}^{a}=(R/2+3\Lambda)e^a$, the $E^{ab}$ component of the Weyl tensor vanishes and hence the on-shell Riemann tensor reads $R^{ab}=C^{ab}-\Lambda e^a\wedge e^b$, having the property ${}^{\hat*}R^{ab}={}^*R^{ab}$. So that the tensor $W^{ab}=R^{ab}+\Lambda e^a\wedge e^b$ we used throughout the paper can be reasonably called the \emph{on-shell Weyl tensor}.

This tensor has another interesting geometric interpretation. The fundamental fields in gravity, say the vielbein and the spin connection, can be assembled into a single Lie algebra-valued connection. For the case of four-dimensional gravity with a non vanishing cosmological constant (the case with vanishing cosmological constant can then be recovered by an Inonu-Wigner contraction) we consider the Lie group $G={\rm SO}\left(3,2\right)$ or $G={\rm SO}\left(4,1\right)$, depending on wheter $\sigma_\perp=\pm 1$ respectively, whose algebra $\mathfrak{g}$ is generated by the standard four-di\-men\-sio\-nal Poincar\'e generators, $P_a$ and $J_{ab}$ with $a,b=0,1,2,3$, with the introduction of a non-commutativity between translations
\begin{eqnarray*}
[P_a,P_b]=-\Lambda J_{ab}\,.
\end{eqnarray*}
Picking a $\mathfrak{g}$-valued connection $\A$, it is natural to interpret its components along generators as $\A=e^aP_a-\frac12\omega^{ab}J_{ab}$, where $e^a$ is the vielbein and $\omega^{ab}$ the spin connection. Its curvature $\F=\dd\A+\A\wedge\A$ can thus be written as $\F=T^aP_a-\frac12W^{ab}J_{ab}$, where $T^a$ is the standard definition for the torsion and $W^{ab}=R^{ab}+\Lambda e^a\wedge e^b$ precisely. So that $W^{ab}$ has a geometric interpretation: it is the component of the curvature of a $\mathfrak{g}$-valued connection along Lorentz transformations.

Within this last context one should pay special attention to the Bianchi identities, since the $G$-covariant exterior derivative is different from the simple Lorentz-covariant one due to the presence of the translations. In particular the Bianchi identity reads $\nabla\F=0$, where $\nabla\F=\dd\F+\A\wedge\F-\F\wedge\A$ with, whose components read
\equ
\nabla\F\Bigl|_{P}&=&{\cal D}T^{a}-W^{a}{}_{b}\wedge e^{b}=0\,,\nonumber\\
\nabla\F\Bigl|_{J}&=&{\cal D}W^{ab}+\Lambda e^{a}\wedge T^{b}-\Lambda e^{b}\wedge T^{a}=0\,,\label{Bianchi.W}
\fequ
where ${\cal D}$ is the Lorentz-covariant part of the full $\nabla$. An interesting fact is that it is not possible to have a configuration with vanishing $W^{ab}$ and non-vanishing torsion $T^{a}$, being the condition $W^{ab}=0$ even more restrictive than $R^{ab}=0$.

\section{Lorentzian Yang-Mills theory in first order formalism}
\label{Lor.YM}
We want to develop the first order formalism for a generic YM theory for some Lie group $G$. Call $\A=\varphi\dd t+\tilde\A$ the $\mathfrak{g}$-valued connection and $\F=\dd t\wedge E+\tilde\F$, with $E_{t}=\tilde\F_{ti}=0$, a $\mathfrak{g}$-valued 2-form which, on-shell, shall give the curvature of the potential $\A$, say $\F=\dd\A+\A\wedge\A$. Pick a manifold $\M$, endowed with a metric structure $g$ providing the standard Hodge dual operator ${}^*$. Therefore we have for the field $\F$
\begin{eqnarray*}
{}^*\F=\dd t\wedge B+\widetilde{{}^*\F}\,,
\end{eqnarray*}
where
\begin{eqnarray*}
B_i=\sqrt{-g}\epsilon_{ijk}\left(g^{jt}E^k+\frac12\tilde\F^{jk}\right)\,,\qquad\widetilde{{}^*\F}_{ij}=\sqrt{-g}\epsilon_{ijk}\left(g^{tt}E^k-g^{tk}E^t+g^{tl}g^{km}\tilde\F_{lm}\right)\,,
\end{eqnarray*}
where $\epsilon_{ijk}=\epsilon_{tijk}$ are the three-dimensional Levi-Civita symbols. It is always possible to choose well-adapted coordinates in order to set $g_{tt}=\sigma_\perp$ and $g_{ti}=0$. In this way the metric on $\M$ can be written as
\begin{eqnarray*}
\dd s^2=\sigma_\perp\dd t^2+h_{ij}(t,\vec x)\dd x^i\dd x^j\,,
\end{eqnarray*}
and hence the dual of $\F$ simplifies to\footnote{Notice that, in this case, the determinant of the four-dimensional metric reduces to $\sqrt{-g}=\sqrt{-\sigma_\perp h}$}
\equ
B={}^{\tilde*}\tilde\F\,,\qquad\widetilde{{}^*\F}=\sigma_\perp{}^{\tilde*}E\,.
\fequ
Picking an Ad-invariant, symmetric, non-degenerate bilinear form $\langle\bullet,\bullet\rangle$ on the algebra, the action shall read
\begin{eqnarray*}
S&=&\int_{\M}-\frac12\langle\F\wedge{}^{*}\F\rangle+\langle\left(\dd\A+\A\wedge\A\right)\wedge{}^{*}\F\rangle\nonumber\\
&=&\int_{\M}\dd t\wedge\left[\langle\dot{\tilde \A}\wedge\widetilde{{}^*\F}\rangle-\frac12\left(\langle E\wedge\widetilde{{}^*\F}\rangle+\langle\tilde\F\wedge B\rangle\right)\right.\nonumber\\
&&\phantom{\int_{\M}\dd t\wedge}\left.\phantom{\frac12}+\langle\left(\tilde\dd\tilde\A+\tilde\A\wedge\tilde\A\right)\wedge B\rangle+\langle\varphi,\tilde\nabla\widetilde{{}^*\F}\rangle\right]-\int_{\M}\dd t\wedge\tilde\dd\langle\varphi,\widetilde{{}^*\F}\rangle\,,
\end{eqnarray*}
where the last term is actually a boundary term. Equivalently, if we performed the transformation to bring the metric in the preferred form, the action would read
\eqn
S&=&-\sigma_\perp\int_{\M}\dd t\wedge\left[-\langle\dot{\tilde \A}\wedge{}^{\tilde *}E\rangle+\frac12\left(\langle E\wedge{}^{\tilde *}E\rangle-\langle{}^{\tilde *}B\wedge B\rangle\right)\right.\nonumber\\
&&\phantom{-\sigma_\perp\int_{\M}\dd t\wedge}\left.\phantom{\frac12}-\sigma_\perp\langle\left(\tilde\dd\tilde\A+\tilde\A\wedge\tilde\A\right)\wedge B\rangle-\langle\varphi,\tilde\nabla{}^{\tilde *}E\rangle\right]+\sigma_\perp\int_{\partial\M}\dd t\wedge\langle\varphi,{}^{\tilde*}E\rangle\,.\label{YM-action}
\feqn
It is easy, at this point, to give some interpretations to the fields. $\varphi$ plays the role of a Lagrange multiplier for the constraint $\tilde\nabla{}^{\tilde *}E=0$, the Gauss law, which is obtained by varying the action with respect to  $\varphi$ itself. The dynamical fields, conjugate to each other, are given by the potential $\tilde\A$ and the electric field $E$, while the magnetic field is some external field. The Lagrange multiplier can be fixed to zero by a gauge transformation, say a certain $g\in G$ such that $\varphi=g^{-1}\dot g$. Hence we are left with a residual gauge symmetry given by group elements $\tilde g\in G$ such that $\dot{\tilde g}=0$. Therefore, within this gauge fixing, the equations of motion read
\eqn
\frac{\delta S}{\delta {}^{\tilde *}E}&=&\dot{\tilde{\A}}-E=0\,,\label{lYM.1}\\
\frac{\delta S}{\delta\tilde\A}&=&-\sigma_\perp\left({}^{\tilde *}E\right)\dot{}+\tilde\nabla B=0\,,\label{lYM.2}\\
\frac{\delta S}{\delta B}&=&\tilde\dd\tilde\A+\tilde\A\wedge\tilde\A+\sigma_\perp{}^{\tilde *}B=0\,,\label{lYM.3}
\feqn
plus the Gauss law
\begin{eqnarray*}
\tilde\nabla{}^{\tilde *}E=0\,.
\end{eqnarray*}
If we want to write them only in terms of the curvature it is easy to see that \eqref{lYM.3} implies the Bianchi identity $\tilde\nabla{}^{\tilde*}B=0$, while combining \eqref{lYM.1} and \eqref{lYM.3} we get $\sigma_\perp\left({}^{\tilde*}B\right)\dot{}+\tilde\nabla E=0$.

It is easy to see that if we define the complex $\mathfrak{g}$-valued 1-form
\equ
{\cal E}\equiv E+iB\,,
\fequ
the equations can be nicely written as
\begin{eqnarray*}
\tilde\nabla{}^{\tilde *}{\cal E}=0\,,\qquad\tilde\nabla{\cal E}-i\sigma_\perp\left({}^{\tilde*}{\cal E}\right)\dot{}=0\,.\label{lYM.complex}
\end{eqnarray*}
The great benefit we acquire is that this form makes explicit the ``global'' duality invariance of the equations of motion
\begin{eqnarray*}
{\cal E}\;\mapsto\;{\cal E}'=e^{i\theta}{\cal E}\,,
\end{eqnarray*}
since the equations are linear and holomorphic in ${\cal E}$.

\section{Example: holography of black holes in AdS$_4$}
As an application of our ideas we consider the holographic description of the standard Schwarzschild AdS$_4$ and also  Taub-NUT-AdS$_4$ black holes with negative cosmological constant. For positive cosmological constant they are still solutions to Einstein's equations but they describe cosmological spacetimes. Our aim is to identify the right initial values -- ``position" and ``velocity" -- which, by the arguments given in the previous section, is equivalent to finding the boundary metric and energy momentum tensor.

We start with the metric
\equ
\label{AdSBH}
\dd s^2=\sigma_\perp\frac{\dd r^2}{V(r)}-\sigma_\perp V(r)\dd\tau^2+r^2\dd\Omega_{\kappa}^{2}\,,
\fequ
that gives the standard  Schwarzschild AdS$_4$ black holes for $\sigma_{\perp}=1$.  The difference with the previous section is the presence of the nontrivial Lapse function $N(r)^2=V(r)^{-1}$ \footnote{See \eqref{esplit} for a definition of the function $N$. Here it has not been gauge-fixed to $1$.} where
\equ
\label{lapse1}
V(r)=\sigma_\perp\kappa-\frac{2M}{r}+\frac{r^2}{\ell^2}\,.
\fequ
The term $\dd\Omega^{2}_{\kappa}$ in \eqref{AdSBH} describes the metric of the horizon which is $S^{2}$, $\mathbb{R}^{2}$ or $H^{2}$ for $\kappa=1, 0, -1$ respectively. Using stereographic projections the horizon metric  can be written in terms of complex coordinates $\{w,\bar w\}$, with $w=x+iy$, as 
\begin{eqnarray*}
\dd\Omega^{2}_{\kappa}=e^{2\gamma}\dd w\dd\bar w\,,\qquad e^{\gamma}=(1+\kappa|w|^{2}/4)^{-1}\,.
\end{eqnarray*}
The vielbein is given by 
\begin{eqnarray*}
e^{0}=V(r)^{-1/2}\dd r\,,\qquad\tilde e^{3}=V(r)^{1/2}\dd\tau\,,\qquad\tilde e^{\bullet}=re^{\gamma}\dd w\,.
\end{eqnarray*}
Solving for the vanishing of the torsion constraints we obtain the electric field
\begin{eqnarray*}
K^3=-\left(\frac{M}{r^2}+\frac{r}{\ell^2}\right)\dd z\,,\qquad K^\bullet=-V(r)^{1/2}e^{\gamma}\dd w\,,
\end{eqnarray*}
and the magnetic field
\begin{eqnarray*}
B^3=-i\left(\partial\gamma\dd w-\bar\partial\gamma\dd\bar w\right)\,,\qquad B^\bullet=0\,.
\end{eqnarray*}
It turns out that the magnetic field is fixed from the boundary, while the electric field depends on the radial coordinate, describing the extrinsic curvature of the metric $\tilde e^{\alpha}$ on the slices $\Sigma_{r}$ at fixed radial coordinate. 

The three-dimensional component of the Weyl tensor reads
\equ
\tilde W^3=-i\sigma_\perp\frac{M}{r^3}\tilde e^\bullet\wedge\tilde e^{\bar\bullet}\,,\qquad\tilde W^\bullet=-i\frac{M}{2r^3}\tilde e^3\wedge\tilde e^\bullet\,. \label{ADSW}
\fequ
and its three-dimensional Hodge dual ${\cal P}^{\alpha}={}^{\tilde*}\tilde W^{\alpha}$ hence reads
\equ
{\cal P}^{3}=-\sigma_\perp\frac{2M}{r^3}\tilde e^3\,,\qquad{\cal P}^\bullet=\sigma_{\perp}\frac{M}{r^3}\tilde e^\bullet\,,\label{ADSW.dual}
\fequ
which is manifestly symmetric (it is actually diagonal) and traceless. 

However, since the metric is not given in the FG form due to the presence of a nontrivial lapse function, we can not directly read from the results above the proper initial data. In general one is not able to compute exactly the diffeomorphism $r=r(t)$, where $t$ is the transverse coordinate bringing the metric into the FG from, nevertheless we present below a general general argument in order to compute the boundary data in some simple cases. Consider  a metric of the form
\equ
\dd s^2=\sigma_\perp N(\rho)^2\dd\rho^2+h_{ij}(\rho,\vec x)\dd x^i\dd x^j\label{BH.gen.noFG}
\fequ
where $N(\rho)=1+\zeta(\rho)$ with $\zeta(\rho)\to0$ as $\rho\to\infty$. For instance this can be achieved in \eqref{AdSBH} by simply defining $r/\ell=e^{\rho/\ell}$. It is clear that if there exists a transformation $t=t(\rho)$ such that
\equ
e^{t/\ell}=e^{\rho/\ell}\left[1+\epsilon(\rho)\right]\,,\label{asympt.transf}
\fequ
with
\equ
\lim_{\rho\to\infty}\epsilon(\rho)=0\,,\label{trasnf.behave}
\fequ
the boundary data can be easily extracted by looking at the leading $\rho\rightarrow\infty$ behavior of the vielbein $\tilde e^\alpha$ and of ${\cal P}^\alpha$. The point now is to understand under which circumstances such a transformation \eqref{asympt.transf} does exist. To bring \eqref{BH.gen.noFG} in the FG form one needs to solve for
\equ
N(\rho)\dd\rho=\dd t\quad \Rightarrow \quad\frac{\zeta(\rho)}\ell=\frac{\epsilon'(\rho)}{1+\epsilon(\rho)}\,.\label{eq.transf.gen}
\fequ
Then, by virtue of \eqref{trasnf.behave} the leading term of \eqref{eq.transf.gen} is given by $\epsilon'=\zeta/\ell$.  Furthermore in order for \eqref{trasnf.behave} to be true, we need to ask for $\lim_{\rho\to\infty}\int\zeta(\rho)$ to be finite.

In our case we have that $\int\zeta(\rho)\propto e^{-2\rho/\ell}$ as $\rho\to\infty$ and hence the boundary data can be easily computed. The boundary vielbein reads
\begin{eqnarray*}
E^{3}=\dd\tau\,,\qquad E^{\bullet}=\ell\,e^{\gamma}\dd w\,,
\end{eqnarray*}
and describes a conformally flat cylinder $\mathbb R\times{\cal B}_{\kappa}$ with base manifold ${\cal B}_{\kappa}$ being  $S^{2},\,\mathbb{R}^{2}$ or $H^{2}$ for $\kappa=1,0,-1$ respectively.  Also, $F^{\alpha}_{[3]}$ is given by
\equ
F^{3}_{[3]}=-\frac{2M\ell^{2}}3E^3\,,\qquad F^{\bullet}_{[3]}=\frac{M\ell^{2}}3E^\bullet\,.\label{init.vel.bh}
\fequ
As a consequence, the vacuum expectation value of the stress tensor of the dual theory simply reads
\equ
\langle T_{33}\rangle_s=\sigma_{\perp}\frac{M\ell}{4\pi G}\,,\qquad\langle T_{\bullet\bar\bullet}\rangle_s=\frac{M\ell}{8\pi G}\,. \label{BHvev}
\fequ
Hence we conclude that these black holes are generated by the evolution along the radial coordinate of the cylinder $\mathbb R\times{\cal B}_{\kappa}$, which is a simple example of a conformally flat manifold, with initial velocity \eqref{init.vel.bh} determined by the black hole mass $M$.

We consider now a generalization of the previous case given by the Taub-NUT-AdS black hole \cite{Astefanesei:2004kn}
\begin{eqnarray*}
ds^2=\sigma_\perp\frac{\dd r^2}{V(r)}-\sigma_\perp V(r)\left(\dd\tau+\sigma\right)^2+\left(r^2+n^2\right)e^{2\gamma}\dd w\dd \bar w \ ,
\end{eqnarray*}
where the lapse function is modified by the NUT charge $n$ to be
\equ
\label{lapse2}
V(r)=\left(\sigma_\perp\kappa+\frac{4n^{2}}{\ell^{2}}\right)\frac{r^2-n^2}{r^2+n^2}-\frac{2Mr}{r^{2}+n^{2}}+\frac{r^{2}+n^{2}}{\ell^{2}}\,.
\fequ
The shift one-form $\sigma$ is given by
\begin{eqnarray*}
\sigma=-i\frac{\sigma_\perp n}{2}e^{\gamma}\left(\bar w\dd w-w\dd\bar w\right) \,. 
\end{eqnarray*}
The metric of the horizon is still given by $e^{\gamma}=(1+\kappa|w|^{2}/4)^{-1}$ with $\kappa=1, 0, -1$. The presence of the shift one-form $\sigma$ introduces a non-staticity in the spacetime since the symmetry under $\tau\mapsto-\tau$ that we had in the previous case is lost. The vielbein $\tilde e^{\alpha}$ on the slices $\Sigma_{r}$ is now given by
\begin{eqnarray*}
\tilde e^{3}=V(r)^{1/2}\left(\dd\tau+\sigma\right)\,,\qquad \tilde e^{\bullet}=(r^{2}+n^{2})^{1/2}e^{\gamma}\dd w\,,
\end{eqnarray*} 
and the electric and magnetic fields read
\begin{eqnarray*}
K^3=-\frac{1}{2}V'(r)V(r)^{-1/2}\tilde e^3\,,\qquad K^\bullet=-\frac{r}{r^2+n^2}V(r)^{1/2}\tilde e^\bullet\,,
\end{eqnarray*}
and
\begin{eqnarray*}
B^3=i\frac\kappa4e^{\gamma}\left(\bar w\dd w-w\dd\bar w\right)-\frac{n}{r^2+n^2}V(r)^{1/2}\tilde e^3\,,\nonumber\\
B^\bullet=\frac{n}{r^2+n^2}V(r)^{1/2}\tilde e^\bullet\,.  
\end{eqnarray*}
The Hodge dual of the three-dimensional spatial component of the on-shell Weyl tensor reads
\equ
{\cal P}^{3}=-\sigma_{\perp}2F(r)\tilde e^{3}\,,\qquad{\cal P}^{\bullet}=\sigma_{\perp}F(r)\tilde e^{\bullet}\label{TNW}
\fequ
where
\begin{eqnarray*}
F(r)=\frac{Mr\left(r^{2}-3n^{2}\right)+n^{2}\left(\sigma_{\perp}\kappa+\frac{4n^{2}}{\ell^{2}}\right)\left(3r^{2}-n^{2}\right)}{(r^{2}+n^{2})^{3}}\,.
\end{eqnarray*}
To extract the proper boundary data we use the general argument above that applies in this case too. Setting as before $r/\ell=e^{\rho/\ell}$ the boundary data are read from the leading terms of the expansions of $\tilde e^{\alpha}$ and ${\cal P}^{\alpha}$. The boundary vielbein is given by
\equ
E^3=\dd\tau +\sigma\ , \qquad E^\bullet=\ell\,e^\gamma \dd w \,,\label{TNUT.bound.viel}
\fequ
which is a sort of a nontrivial line bundle with base manifold the same ${\cal B}_{\kappa}$ as in the previous black hole case. Despite being conformally flat the boundary manifold is rather nontrivial due to the presence of the NUT charge. Its most striking feature is the presence, for some values of the the NUT parameter, of closed timelike curves (CTCs) \cite{Astefanesei:2004kn}. In order to see this consider spherical coordinates on the base ${\cal B}_{\kappa}$ and hence we have
\begin{eqnarray*}
\dd\Omega_{\kappa}^{2}=\dd\theta^{2}+f_{\kappa}(\theta)^{2}\dd\phi^{2}\,,\qquad\sigma=4nf_{\kappa}(\theta/2)^{2}\,,
\end{eqnarray*}
where
\begin{eqnarray*}
f_{\kappa}(\theta)=\left\{
\begin{array}{lcl}
\sin\theta&&{\rm if\;}\kappa=1\\
\theta&&{\rm if\;}\kappa=0\\
\sinh\theta&&{\rm if\;}\kappa=-1\\
\end{array}\right.
\end{eqnarray*}
In all three cases $\phi$ is an angular coordinate and we can see that
\begin{eqnarray*}
g_{\phi\phi}=f_{\kappa}(\theta)^{2}-16\frac{n^{2}}{\ell^{2}}f_{\kappa}(\theta/2)^{4}
\end{eqnarray*}
becomes {\it negative} for certain values of $\theta$ and $n/\ell$. For these values the vector $\partial_{\phi}$, which generates closed curves parameterized by the angle $\phi$, becomes {\it timelike}. In particular for $\kappa=-1$ this three-dimensional spacetime is a 3D slice of the G\"odel spacetime \cite{Astefanesei:2004kn}, but it encodes all the features of such a manifold. Therefore these bulk metrics could be used to study three-dimensional  quantum field theories (at least endowed with conformal invariance) on spacetimes with causal pathologies. For $F^{\alpha}_{[3]}$ we obtain the following results
\equ
F^{3}_{[3]}=-\frac{2M\ell^{2}}3E^3\,,\qquad F^{\bullet}_{[3]}=\frac{M\ell^{2}}3E^\bullet\,,\label{init.vel.bh.nut}
\fequ
and hence the tangent components of the boundary energy-momentum tensor are the same as before \eqref{BHvev}. What changes here is the background where the NUT charge introduces a nontriviality, nevertheless keeping it conformally flat. From the initial value problem point of view, these spacetimes are generated by the evolution of the 3D metric described by \eqref{TNUT.bound.viel} with the same initial velocity as before.


\begin{thebibliography}{99}

\bibitem{Sachdev}
  C.~P.~Herzog, P.~Kovtun, S.~Sachdev and D.~T.~Son,
  ``Quantum critical transport, duality, and M-theory,''
  Phys.\ Rev.\  D {\bf 75} (2007) 085020
  [arXiv:hep-th/0701036].

\bibitem{HP}
  S.~A.~Hartnoll and P.~Kovtun,
  ``Hall conductivity from dyonic black holes,''
  Phys.\ Rev.\  D {\bf 76} (2007) 066001
  [arXiv:0704.1160 [hep-th]].

\bibitem{HH}
  S.~A.~Hartnoll and C.~P.~Herzog,
  ``Ohm's Law at strong coupling: S duality and the cyclotron resonance,''
  Phys.\ Rev.\  D {\bf 76} (2007) 106012
  [arXiv:0706.3228 [hep-th]].
  
 \bibitem{Horowitz}
  S.~A.~Hartnoll, C.~P.~Herzog and G.~T.~Horowitz,
  ``Building an AdS/CFT superconductor,''
  arXiv:0803.3295 [hep-th].

\bibitem{Gubser}
  S.~S.~Gubser and S.~S.~Pufu,
  ``The gravity dual of a p-wave superconductor,''
  arXiv:0805.2960 [hep-th].

\bibitem{BL1}
  J.~Bagger and N.~Lambert,
  ``Modeling multiple M2's,''
  Phys.\ Rev.\  D {\bf 75} (2007) 045020
  [arXiv:hep-th/0611108].
  
\bibitem{BL2}
  J.~Bagger and N.~Lambert,
  ``Gauge Symmetry and Supersymmetry of Multiple M2-Branes,''
  Phys.\ Rev.\  D {\bf 77} (2008) 065008
  [arXiv:0711.0955 [hep-th]].
  
\bibitem{BL3}
  J.~Bagger and N.~Lambert,
  ``Comments On Multiple M2-branes,''
  JHEP {\bf 0802} (2008) 105
  [arXiv:0712.3738 [hep-th]].

\bibitem{Gustavsson}
  A.~Gustavsson,
  ``Algebraic structures on parallel M2-branes,''
  arXiv:0709.1260 [hep-th].

\bibitem{ABJM}
  O.~Aharony, O.~Bergman, D.~L.~Jafferis and J.~Maldacena,
  ``N=6 superconformal Chern-Simons-matter theories, M2-branes and their gravity duals,''
  arXiv:0806.1218 [hep-th].

\bibitem{BD}
  C.~P.~Burgess and B.~P.~Dolan,
  ``Particle-vortex duality and the modular group: Applications to the  quantum
  Hall effect and other 2-D systems,''
  arXiv:hep-th/0010246.
  
\bibitem{Witten}
  E.~Witten,
  ``SL(2,Z) action on three-dimensional conformal field theories with Abelian symmetry,''
  arXiv:hep-th/0307041.

\bibitem{LP1}
  R.~G.~Leigh and A.~C.~Petkou,
  ``SL(2,Z) action on three-dimensional CFTs and holography,''
  JHEP {\bf 0312} (2003) 020
  [arXiv:hep-th/0309177].
  
\bibitem{dHP1}
  S.~de Haro and A.~C.~Petkou,
  ``Instantons and conformal holography,''
  JHEP {\bf 0612} (2006) 076
  [arXiv:hep-th/0606276].
  
  
\bibitem{dHG}
  S.~de Haro and P.~Gao,
  ``Electric-magnetic Duality and Deformations of Three-Dimensional CFT's,''
  Phys.\ Rev.\  D {\bf 76} (2007) 106008
  [arXiv:hep-th/0701144].
  
  
\bibitem{dHPP}
  S.~de Haro, I.~Papadimitriou and A.~C.~Petkou,
  ``Conformally coupled scalars, instantons and vacuum instability in AdS(4),''
  Phys.\ Rev.\ Lett.\  {\bf 98} (2007) 231601
  [arXiv:hep-th/0611315].
  
\bibitem{dHPConf}
  S.~de Haro and A.~C.~Petkou,
  ``Holographic Aspects of Electric-Magnetic Dualities,''
  J.\ Phys.\ Conf.\ Ser.\  {\bf 110} (2008) 102003
  [arXiv:0710.0965 [hep-th]].
  
  
\bibitem{Marolf}
  G.~Compere and D.~Marolf,
  ``Setting the boundary free in AdS/CFT,''
  arXiv:0805.1902 [hep-th].
  
  
  \bibitem{MPT2}
  "Gravity in the 3+1-Split Formalism II: Self-Duality and the Emergence of Gravitational Chern-Simons in the Boundary", D. Mansi, A. C. Petkou and G. Tagliabue, to appear.
  
\bibitem{LP2}
  R.~G.~Leigh and A.~C.~Petkou,
  ``Gravitational Duality Transformations on (A)dS4,''
  JHEP {\bf 0711} (2007) 079
  [arXiv:0704.0531 [hep-th]].

\bibitem{ADM}
  R.~Arnowitt, S.~Deser and C.~W.~Misner,
  ``The dynamics of general relativity,''
  arXiv:gr-qc/0405109.

\bibitem{BK}
  V.~Balasubramanian and P.~Kraus,
  ``A stress tensor for anti-de Sitter gravity,''
  Commun.\ Math.\ Phys.\  {\bf 208}, 413 (1999)
  [arXiv:hep-th/9902121].


\bibitem{dHSS}
  S.~de Haro, S.~N.~Solodukhin and K.~Skenderis,
  ``Holographic reconstruction of spacetime and renormalization in the  AdS/CFT
  correspondence,''
  Commun.\ Math.\ Phys.\  {\bf 217} (2001) 595
  [arXiv:hep-th/0002230].

\bibitem{Zanelli}
  R.~Aros, M.~Contreras, R.~Olea, R.~Troncoso and J.~Zanelli,
  ``Conserved charges for gravity with locally AdS asymptotics,''
  Phys.\ Rev.\ Lett.\  {\bf 84} (2000) 1647
  [arXiv:gr-qc/9909015].

\bibitem{Mora:2004kb}
  P.~Mora, R.~Olea, R.~Troncoso and J.~Zanelli,
  ``Finite action principle for Chern-Simons AdS gravity,''
  JHEP {\bf 0406}, 036 (2004)
  [arXiv:hep-th/0405267].

\bibitem{Olea}
  R.~Olea,
  ``Regularization of odd-dimensional AdS gravity: Kounterterms,''
  JHEP {\bf 0704} (2007) 073
  [arXiv:hep-th/0610230].


\bibitem{Gibbons:1976ue}
  G.~W.~Gibbons and S.~W.~Hawking,
  ``Action Integrals And Partition Functions In Quantum Gravity,''
  Phys.\ Rev.\  D {\bf 15} (1977) 2752.

\bibitem{LNP}
"Torsion vortices and Holography", R. G. Leigh, N. N. Hoang and A. C. Petkou, to appear

\bibitem{Wald:GR}
  Robert~M. Wald,
  ``General Relativity'', 
  The University of Chicago Press, 1984.



\bibitem{IshiWald1}
  A.~Ishibashi and R.~M.~Wald,
  ``Dynamics in non-globally-hyperbolic static spacetimes. II: General
  analysis of prescriptions for dynamics,''
  Class.\ Quant.\ Grav.\  {\bf 20} (2003) 3815
  [arXiv:gr-qc/0305012].


\bibitem{IshiWald2}
  A.~Ishibashi and R.~M.~Wald,
  ``Dynamics in non-globally hyperbolic static spacetimes. III: anti-de  Sitter spacetime,''
  Class.\ Quant.\ Grav.\  {\bf 21} (2004) 2981
  [arXiv:hep-th/0402184].


\bibitem{FG}
  C.~Fefferman and C.~R.~Graham,
  ``Conformal invariants,''
  in {\it \'Elie Cartan et les Math\'ematiques d'Aujourd'hui}, Ast\'erisque,
  (1985), 95-116. 

\bibitem{fefferman-2007}
  C.~Fefferman and C.~R.~Graham,
  ``The Ambient Metric'',
  [arXiv:0710.0919].

\bibitem{Skenderis:2002wp}
  K.~Skenderis,
  ``Lecture notes on holographic renormalization,''
  Class.\ Quant.\ Grav.\  {\bf 19}, 5849 (2002)
  [arXiv:hep-th/0209067].
  
\bibitem{WittenAdSCFT}
  E.~Witten,
  ``Anti-de Sitter space and holography,''
  Adv.\ Theor.\ Math.\ Phys.\  {\bf 2} (1998) 253
  [arXiv:hep-th/9802150].

\bibitem{graham-2004}
  C.~Robin Graham and Kengo Hirachi,
  ``The Ambient Obstruction Tensor and Q-Curvature'',
  [arXiv:math/0405068].

\bibitem{MacDowell:1977jt}
  S.~W.~MacDowell and F.~Mansouri,
  ``Unified Geometric Theory Of Gravity And Supergravity,''
  Phys.\ Rev.\ Lett.\  {\bf 38} (1977) 739
  [Erratum-ibid.\  {\bf 38} (1977) 1376].

\bibitem{Anabalon:2006fj}
  A.~Anabalon, S.~Willison and J.~Zanelli,
  ``General relativity from a gauged WZW term,''
  Phys.\ Rev.\  D {\bf 75}, 024009 (2007)
  [arXiv:hep-th/0610136].

\bibitem{Anabalon:2007dr}
  A.~Anabalon, S.~Willison and J.~Zanelli,
  ``The Universe as a topological defect,''
  Phys.\ Rev.\  D {\bf 77}, 044019 (2008)
  [arXiv:hep-th/0702192].

\bibitem{Anabalon:2008hi}
  A.~Anabalon,
  ``Some considerations on the Mac Dowell-Mansouri action,''
  JHEP {\bf 0806}, 069 (2008)
  [arXiv:0805.3558 [hep-th]].

\bibitem{Wise:2006sm}
  D.~K.~Wise,
  ``MacDowell-Mansouri gravity and Cartan geometry,''
  arXiv:gr-qc/0611154.
  
\bibitem{Astefanesei:2004kn}
  D.~Astefanesei, R.~B.~Mann and E.~Radu,
  ``Nut charged space-times and closed timelike curves on the boundary,''
  JHEP {\bf 0501} (2005) 049
  [arXiv:hep-th/0407110].
  


\bibitem{Smolin1}
  L.~Smolin,
  ``Linking topological quantum field theory and nonperturbative quantum gravity,''
  J.\ Math.\ Phys.\  {\bf 36} (1995) 6417
  [arXiv:gr-qc/9505028].

\bibitem{Smolin2}
  L.~Smolin,
  ``A holographic formulation of quantum general relativity,''
  Phys.\ Rev.\  D {\bf 61} (2000) 084007
  [arXiv:hep-th/9808191].

\bibitem{Smolin3}
  L.~Smolin,
  ``Quantum gravity with a positive cosmological constant,''
  arXiv:hep-th/0209079.






%


\end{thebibliography}
\end{document}